\documentclass{article}
\usepackage{amsfonts,amssymb, amsmath}
\usepackage[english]{babel}

\DeclareMathAlphabet{\mathpzc}{OT1}{pzc}{m}{it}

\def\nn{\nonumber }
\def\bq{ \begin{equation} }
\def\eq{ \end{equation} }
\def\ben{ \begin{eqnarray} }
\def\en{ \end{eqnarray} }

\def\e{{\rm e}}
\def\ii{{\rm i}}

\newtheorem{prop}{Proposition}

\newtheorem{re}{Remark}

\begin{document}


\title{On bi-hamiltonian geometry of some integrable systems on the sphere with cubic integral of motion. }
\author{A V Vershilov and A V Tsiganov \\
\it\small
St.Petersburg State University, St.Petersburg, Russia\\
\it\small e--mail: alexander.vershilov@gmail.com and tsiganov@mph.phys.spbu.ru}

\date{}
\maketitle

\begin{abstract}
We obtain  bi-Hamiltonian structure for a family of integrable systems on the sphere ${\mathcal S}^2$
with an additional integral of third order in momenta. These results are applied to the
Goryachev system and Goryachev-Chaplygin top for which we give an explicit procedure to find the separated coordinates and the separated relations.
\end{abstract}


\vskip0.1truecm

\section{Introduction}
\setcounter{equation}{0}
We address the problem of the separation of variables for the Hamilton-Jacobi equation within the theoretical scheme of bi-hamiltonian geometry. We want to learn to calculate bi-hamiltonian structure for a given integrable system on the Poisson manifold $M$ with the Poisson bivector $P$ and the Casimir functions $C_a$. The separation variables are naturally associated with this bi-hamiltonian  structure of $M$ itself \cite{fp02}.

According to \cite{ts07c,ts08,ts08b} we will suppose that desired second Poisson bivector is
the Lie derivative of $P$ along some unknown Liouville vector field $X$ on $M$
\bq\label{co-b}
P'=\mathcal L_X(P)
\eq
which has to satisfy to the equation
\bq\label{m-eq2}
[P',P']=[\mathcal L_X(P),\mathcal L_X(P)]=0,\quad\Leftrightarrow\quad\
[\mathcal L_X^2(P),P]=0,
\eq
with respect to the Schouten bracket $[.,.]$. By definition (\ref{co-b}) bivector $P'$ is compatible with a given bivector $P$, i.e. $[P,P']=0$. In geometry such bivector $P'$ is said to be
the 2-coboundary associated with the Liouville vector field  $X$
in the Poisson-Lichnerowicz cohomology defined by $P$ on $M$.

From all the solutions $X$ of the equation (\ref{m-eq2})
we have to choose partial solutions $X$ such that
\bq\label{bi-int}
\{H_i,H_j\}'=0,\qquad  i,j=1,\ldots,n,
\eq
where $H_j$ are the given integrals of motion and $\{.,.\}'$ is the  Poisson bracket
associated with the  Poisson bivector $P'$ (\ref{co-b}).

Obviously enough, in their full generality the system of equations (\ref{m-eq2}-\ref{bi-int}) is too difficult to be solved because it has infinitely many solutions labeled by  different separated coordinates, see  \cite{ts08b}. In order to get particular solutions we will use addition assumptions
\bq\label{m-eq3}
P'dC_a=0,\qquad a=1,\ldots,k,
\eq
and some special ans\"{a}tze for the Liouville vector field $X$.

Equations (\ref{m-eq3}) say that we are looking for the Poisson bivector $P'$, which has
the same foliation by symplectic leaves as $P$. In general
bivector $P'$ could have some more Casimirs, so that their symplectic leaf could be smaller, but
in the separation of variables method we have to consider equivalent foliations only \cite{ts07c,ts08,ts08b}.

\section{A family of integrable systems on the sphere}
\setcounter{equation}{0}
A Hamiltonian system is called natural if its Hamiltonian is the sum of a positive-definite
kinetic energy and a potential. Natural Hamiltonian systems on cotangent bundles of closed surfaces admitting integrals polynomial in momenta are especially interesting \cite{darb}. In this Section we define some family of natural Hamiltonian systems on $T^*{\mathcal S}^2$ with the quadratic  and cubic   in the momenta integrals of motion.

Let two vectors $J=(J_1,J_2,J_3)$ and $ x=(x_1,x_2,x_3)$ are coordinates on
the Euclidean algebra $e^*(3)$ with the Lie-Poisson bracket
\begin{equation}\label{e3}
\,\qquad \bigl\{J_i\,,J_j\,\bigr\}=\varepsilon_{ijk}J_k\,, \qquad
\bigl\{J_i\,,x_j\,\bigr\}=\varepsilon_{ijk}x_k \,, \qquad
\bigl\{x_i\,,x_j\,\bigr\}=0\,,
\end{equation}
where $\varepsilon_{ijk}$ is the totally skew-symmetric tensor.
This bracket has two Casimir functions
\bq \label{caz-e3}
C_1=|x|^2\equiv\sum_{k=1}^3 x_k^2, \qquad C_2= (x,J)\equiv\sum_{k=1}^3 x_kJ_k .
\eq
Fixing their values one gets a generic symplectic leaf of $e^*(3)$
\[
{\mathcal O}_{AB}:\qquad \{{x}\,, {J}\,:~C_1=A,~~
C_2=B\}\,,
\]
which is a four-dimensional symplectic manifold.

At $C_2=0$ this symplectic manifold
is  equivalent to cotangent bundle $T^*{\mathcal S}^2$
of the sphere
\[\mathcal S^2=\{x\in \mathbb R^3,\qquad |x|=A\}.\]
For the Liouville integrability of the equations of
motion  it is enough to find one additional integral of
motion $H_2$, which is functionally independent of the Hamiltonian
$H_1$ and the Casimir functions.

In this $e^*(3)$ coordinates the family of integrable systems on the sphere
is defined by the following quadratic Hamiltonian
\bq\label{H}
H_1=J_1^2+J_2^2+\Bigl(3\alpha^2+f(x_3)\Bigr)J_3^2+m(x_3)x_1+g(x_3)\,,
\eq
and cubic additional integral of motion
\bq\label{K}
H_2= -2 \alpha J_3 \Bigl(-\alpha^2 J_3^2+J_1^2+J_2^2+f(x_3)
J_3^2 +g(x_3)\Bigr)-n(x_3) J_1-\ell(x_3) x_1 J_3\,.
\eq
Here
\ben
g(x_3)=\dfrac{a^2b}{n(x_3)^2},\qquad
m(x_3)=-\dfrac{n'(x_3)}{\alpha},
\qquad \ell(x_3)= \dfrac{n(x_3)n''(x_3)}{n'(x_3)}, \nn\\
\label{fgml}\\
f(x_3)=
1-3\alpha^2-\alpha\,\dfrac{3x_3m(x_3)-2(A^2-x_3^2)m'(x_3)}{n(x_3)}
+\dfrac{x_3\ell(x_3) -(A^2-x_3^2)\ell'(x_3)}{n(x_3)}\,,\nn
\en
and
\bq\label{anz}
n(x_3)=-a(x_3+c)^\beta\,,\qquad a,b,c\in\mathbb R\,.
\eq
At $C_2=0$ integrals of motion $H_1$ (\ref{H}) and $H_2$
(\ref{K}) are in the involution with respect to the brackets
(\ref{e3}) in the  following five cases:
\bq\label{msol}
\begin{array}{llll}
 1. &\alpha=\beta=1,\quad & c=0,\qquad& n(x_3)=-ax_3,\\
 \\
 2. &\alpha=\beta=\dfrac{1}{3},\quad&c=0,\qquad &n(x_3)=-ax_3^{1/3}, \\
 \\
 3. &\alpha=\beta=\dfrac16,\quad &c=A, \qquad &n(x_3)=-a(x_3+A)^{1/6},\\
 \\
 4. &\alpha=\beta=\dfrac12,\quad & c\in \mathbb
 R,\qquad
  &n(x_3)=-a(x_3+c)^{1/2}\,,\\
 \\
 5. &\alpha=\beta =\dfrac14,\quad &c=A,\qquad
 &n(x_3)=-a(x_3+A)^{1/4},.
\end{array}
\eq
The corresponding Hamiltonian (\ref{H}) depends on parameters $a,b$ and $c$:
\ben
H_1^{(1)}&=&J_1^2+J_2^2+4J_3^2+ax_1+\dfrac{b}{x_3^2}\,,\nn\\
\nn\\
H_1^{(2)}&=&J_1^2+J_2^2+\dfrac{4}{3}J_3^2+\dfrac{ax_1}{x_3^{2/3}}+\dfrac{b}{x_3^{2/3}}\label{H5}\\
\nn\\
H_1^{(3)}&=&J_1^2+J_2^2+\left(\dfrac{7}{12}+\dfrac{x_3}{2(x_3+A)}\right)
J_3^2
+\dfrac{ax_1}{(x_3+A)^{5/6}}+\dfrac{b}{(x_3+A)^{1/3}}\,,\nn\\
\nn\\
H_1^{(4)}&=&J_1^2+J_2^2+\left(1+\dfrac{x_3}{x_3+c}-\dfrac{x_3^2-c^2}{4(x_3+c)^2}\right)
J_3^2+\dfrac{ax_1}{(x_3+c)^{1/2}}+\dfrac{b}{x_3+c}\,,\nn\\
\nn\\
H_1^{(5)}&=&J_1^2+J_2^2+\left(\dfrac{13}{16}+\dfrac{3x_3}{8(x_3+A)}\right)
J_3^2
+\dfrac{ax_1}{(x_3+A)^{3/4}}+\dfrac{b}{(x_3+A)^{1/2}}\,.\nn
\en
The Hamilton function $H_1^{(1)}$ describes the well-studied Goryachev-Chaplygin
top \cite{ch48}. In contrast with this case the other systems have not at all physical meaning, however they may be interesting as some mathematical toys.
The second integrable system with the Hamiltonian $H_1^{(2)}$ was found by Goryachev \cite{gor16}. The
Hamilton function $H_1^{(4)}$  was studied by Dullin and Matveev
\cite{dull04}. The third and fifth integrable systems with
Hamiltonians $H_1^{(3)}$ and $H_1^{(5)}$ have been found in \cite{ts05}.

\section{Construction of the Liouville vector field}
\setcounter{equation}{0}
In this Section we describe how to obtain  desired Liouville vector field $X$ and the corresponding second Poisson bivector $P'$ for our family of integrable systems. Remind, that smooth manifold $M$ endowed with a pair of compatible Poisson bivectors $P$ and $P'$ is said to be $\omega N$ manifold if one of the Poisson brackets is non degenerate \cite{fp02}.

Without loss of generality we can always put $|x|=A=1$ by using scaling transformation $x_j\to A^{-1} x_j$
and consider the unit sphere ${\mathcal S}^2$ only. On its cotangent bundle $T^*{\mathcal S}^2$ we introduce the following  coordinates
\[
\phi =\arctan\left(\dfrac{x_1}{x_2}\right),\qquad u = x_3,\qquad
p_\phi=-J_3,\qquad p_u =\dfrac{J_1x_2-x_1J_2}{x_1^2+x_2^2},
\]
so that
\[\begin{array}{ll}
J_1=\dfrac{u}{\sqrt{1-u^2}}\sin(\phi)\,p_\phi+\sqrt{1-u^2}\cos(\phi)\,p_u,\quad &
x_1 = \sqrt{1-u^2}\sin(\phi),\\
\\
J_2=\dfrac{u}{\sqrt{1-u^2}}\cos(\phi)\,p_\phi-\sqrt{1-u^2}\sin(\phi)\,p_u,\quad&
x_2 = \sqrt{1-u^2}\cos(\phi),\\
\\
J_3 = -p_\phi,\quad x_3 = u.
\end{array}
\]
In these coordinates $u,\phi,p_u,p_\phi$ initial Poisson tensor $P$ associated with the brackets (\ref{e3}) becomes canonical tensor on 2-dimensional symplectic manifold
\bq\label{can-Poi}
 P=\left(%
\begin{array}{cccc}
 0&0&1&0\\
 0&0&0&1\\
-1&0&0&0\\
0&-1&0&0
\end{array}%
\right),
\eq
which is non degenerate.

Calculations in these coordinates are faster than calculations in the standard spherical coordinates $\phi,\theta=\arccos u$. Of course, using these coordinates on symplectic leaf of $e^*(3)$ we implicitly accepted assumptions (\ref{m-eq3}) for the initial Poisson manifold $e^*(3)$ with the Casimir functions (\ref{caz-e3}).

The Hamiltonian (\ref{H}) in these coordinates looks like
\ben
H_1&=&(1-u^2)p_u^2+\left(3\alpha^2+f(u)+\dfrac{u^2}{1-u^2}\right)p_\phi^2+\sqrt{1-u^2}\sin(\phi)\,m(u)+g(u)
\nonumber\\
&=&\mathrm{g}_u\,p_u^2+\mathrm{g}_\phi\,p_\phi^2+V(u,\phi)\,,\label{g-phi}
\en
where
\[
\mathrm{g}_u=(1-u^2),\qquad \mathrm{g}_\phi=3\alpha^2+f(u)+\dfrac{u^2}{1-u^2}\,.
\]

According to \cite{ts06} in order to solve equations (\ref{m-eq2}-\ref{bi-int}) we will use polynomial in momenta \textit{ans\"{a}tze} for the components of the Liouville vector field $X=\sum X^j\,\partial_j$:
\bq\label{anz}
X^j=\sum_{k=0}^N\sum_{m=0}^k y^j_{km}(u,\phi) p_u^{k-m} p_\phi^m.
\eq
For all the systems we put $N=2$, it means that $X^j$ will be generic second order polynomials in momenta $p_u,p_\phi$ with coefficients $y^j_{km}(u,\phi)$ depending on variables $u$ and $\phi$.

Substituting this ans{a}tz (\ref{anz}) into the equations (\ref{m-eq2}-\ref{bi-int}) and demanding that
all the coefficients at powers of $p_u$ and $p_\phi$ vanish one gets the over determined system of 60 algebro-differential equations on the 24 functions $y^j_{km}(u,\phi)$ which can be easily solved in the modern computer algebra systems.

Below we discuss all the obtained solutions.

\subsection{Particular solution}

For the Goryachev-Chaplygin top \cite{ch48} with integral of motion $H^{(1)}_{1,2}$ one get the following
\begin{prop}
At the first case $\alpha=\beta=1$ and $n(x_3)=ax_3$ there is one  particular solution depending on coordinate $u$ only:
\bq\label{p-sol}
P'=\left(
     \begin{array}{cccc}
       0 & u & \dfrac{u\, p_\phi}{1-u^2}& (1-u^2)p_u \\
       \\
       * & 0 & \dfrac{(1-2u^2)\, p_u}{1-u^2} & \dfrac{(2-u^2)\, p_\phi}{1-u^2} \\
       \\
       * & * & 0 & up_u^2+\dfrac{up_\phi^2}{1-u^2}+\dfrac{b}{u^3} \\
       \\
       * & * & * & 0 \\
     \end{array}
   \right)\,.
\eq
The entries of $P'$ are real functions on initial variables.
\end{prop}
By construction this Poisson bivector compatible with $P$ (\ref{can-Poi})
and functions $H_{1,2}$ are in bi-involution with respect to the corresponding Poisson brackets.
So that the phase space $T^*{\mathcal S}^2$ becomes semisimple $\omega N$ manifold and the foliation defined by $H_{1,2}$ is separable in the Darboux-Nijenhuis variables  \cite{fp02}.

The variables of separation $q_{1,2}$  (the  Darboux-Nijenhuis variables) are  the eigenvalues of the recursion operator $N=P'P^{-1}$. They are simply roots of the following minimal characteristic polynomial of $N$
\ben
\mathcal A(\lambda)=\Bigr(\mathrm{det}\,(N-\lambda\mathrm{I})\Bigl)^{1/2}&=&\lambda^2-2p_\phi\lambda-(1-u^2)p_u^2-\dfrac{u^2p_\phi^2}{1-u^2}-\dfrac{b}{u^2}=\nn\\
&=&\lambda^2+2J_3\lambda-J_1^2-J_2^2-\dfrac{b}{x_3^2}\,.
\en
At $b=0$ these variables of separation have been found by Chaplygin in  \cite{ch48}.
In initial $e^*(3)$ variables this bivector
\bq\label{sec-e3}
P'=\left(\begin{matrix}
0& -x_3^2& x_3 x_2& - x_2J_1& -x_2 J_2& x_3 J_2-2x_2J_3\\
*& 0& -x_3 x_1& x_1J_1& x_1J_2 & 2x_1 J_3 -x_3 J_1\\
*& *& 0& 0& 0& -x_1J_2+x_2J_1\\
*& *& *& 0& -J_1^2-J_2^2& -J_3 J_2\\
*& *& *& *& 0& J_1 J_3\\
*& *& *& *& *& 0
\end{matrix}\right)
\eq
has been found in \cite{ts07a}. It is easy to prove that  $P'$ has the same foliations by symplectic leaves as  $P$ at $C_2=0$.

\subsection{Generic solution}
Using quadratic in momenta ans{a}tz (\ref{anz}) one got  the generic solution depending on a pair of parameters $\mu$ and $\nu$ for all five cases of integrable systems (\ref{H5}).

\begin{prop}
For integrable system (\ref{H}-\ref{K}) equations (\ref{m-eq2}-\ref{bi-int}) have the following  solution
\bq\label{p-mn}
P_{\mu\nu}'=\left(
     \begin{array}{cccc}
       0 & \mathcal X & \mathcal Y+4\,\ii \alpha\mu p_\phi& 0 \\
       \\
       * & 0 & \mathcal Z & 4\,\ii \alpha\mu p_\phi \\
       \\
       * & * & 0 & \mathcal X\dfrac{\,\ii m(u) \e^{\,\ii \phi}}{2\sqrt{1-u^2}}
 \\
       \\
       * & * & * & 0 \\
     \end{array}
   \right)\,,
\eq
where
\ben
\mathcal X&=&-2\,\ii\,\mu(u+d)+\dfrac{\nu}{n(u)m(u)},\qquad\qquad \ii=\sqrt{-1},\nn\\
\mathcal Y&=&\mathcal X\left[\,\ii p_u+
\left(\dfrac{3\alpha-1}{u+c}-\dfrac{u}{1-u^2}\right)\,p_\phi\right],\nn\\
\nn\\
\mathcal Z&=&\dfrac{\Bigl(-2\,\ii\, \mu\,a^2(u+d)^{2\beta}-\nu\Bigr)}{n(u)^2}\Bigl[
       \Bigl(\alpha-\dfrac{u(u+c)}{(1-u^2)}\Bigr)p_u
-\dfrac{\ii\, (u+c)\mathrm{g}_\phi}{(1-u^2)} p_\phi\Bigr],\nn
\en
Here $\mathrm g_\phi$ is a component of metric (\ref{g-phi}) and
\begin{itemize}
  \item $d=0$ at cases 1,2,4,
  \item $d=A$ at cases 3,5.
\end{itemize}
The complex conjugated bivector
$\overline{P^{\prime}}_{\mu\nu}$ is another solution of the same equations (\ref{m-eq2}-\ref{bi-int}).
\end{prop}

In contrast with the particular solution (\ref{p-sol}) entries of this Poisson bivector are complex functions on initial variables and $P_{\mu\nu}'$ depends on parameter $a$ instead of parameter $b$.

As above this Poisson bivector is compatible with $P$ (\ref{can-Poi}) and functions $H_{1,2}$ are in bi-involution with respect to the corresponding Poisson brackets.
So, the phase space $T^*{\mathcal S}^2$ becomes semisimple $\omega N$ manifold and the foliation defined by $H_{1,2}$ is separable in the Darboux-Nijenhuis variables, which are the eigenvalues of the recursion operator $N=P'P^{-1}$ \cite{fp02}.

Summing up, we have found particular and generic solutions of the equations (\ref{m-eq2}-\ref{bi-int}) for five integrable systems (\ref{H5}) on the sphere  with cubic integrals of motion. An important application of this result is the separation of variables for these systems.

\section{Separation of variables}
\setcounter{equation}{0}
In this section we consider new separated variables and separated relations for the Goryachev system \cite{gor16}
and for the Goryachev-Chaplygin top \cite{ch48} in details.

\subsection{The Goryachev system}

In this case
\[\alpha=\beta=\dfrac{1}{3},\quad n=-ax_3^{1/3},\quad
m=\dfrac{a}{x_3^{2/3}},\quad
\ell=\dfrac{2 a}{3 x_3^{2/3}},\quad
g=\dfrac{b}{x_3^{2/3}},\quad
f=1,
\]
so that
\[H_1^{(2)}=J_1^2+J_2^2+\dfrac{4}{3}\,J_3^2+\dfrac{ax_1}{x_3^{2/3}}+\dfrac{b}{x_3^{2/3}}\]
and
\[
H_2^{(2)}= -\dfrac{2}{3}\,J_3\,\left(J_1^2+J_2^2+\dfrac{8}{9}\,J_3^2+\dfrac{b}{x_3^{2/3}}\right)
+ax_3^{1/3}J_1-\dfrac{2ax_1J_3}{3x_3^{2/3}}\,.
\]
If we put $\mu=0$ and $\nu=-a^2$ in $P_{\mu\nu}'$ (\ref{p-mn}) one gets the following second bivector on $T^*{\mathcal S}^2$:
\[
P'=\left(
     \begin{array}{cccc}
       0 & u^{1/3} & \,\ii\, u^{1/3}p_u-\dfrac{u^{4/3}}{1-u^2}p_\phi & 0 \\
       \\
       * & 0 & \dfrac{1-4u^2}{3u^{2/3}(1-u^2)}\,p_u +\dfrac{\,\ii\, u^{1/3}(4-u^2)}{3(1-u^2)^2}\,p_\phi&  0\\
       \\
       * & * & 0 & \dfrac{\,\ii\,  a\,\e^{\,\ii\, \phi}}{2u^{1/3}\sqrt{1-u^2}} \\
       \\
       * & * & * & 0 \\
     \end{array}
   \right)
\]
The variables of separation $q_{1,2}$  (the  Darboux-Nijenhuis variables) are  eigenvalues of the recursion operator $N=P'P^{-1}$, which are roots of the following polynomial
\bq
\mathcal A(\lambda)=(\lambda-q_1)(\lambda-q_2)=
\lambda^2+u^{1/3}\left(\dfrac{up_\phi}{1-u^2}-\ii\,p_u\right)\lambda-\dfrac{\ii\,a\,\e^{\ii\,\phi}}{2\sqrt{1-u^2}}\,.
\eq

In the initial $e^*(3)$ variables the second Poisson brackets look like
\ben
\{x_i,x_j\}'&=&\varepsilon_{ijk}x_kx_3^{1/3},\qquad\qquad \{x_j,J_3\}'=0,\nn\\
\{x_1,J_1\}'&=&\dfrac{x_2J_1}{3x_3^{2/3}}-\dfrac{x_3^{4/3}J_2}{x_1+\ii\, x_2}
+\dfrac{4x_3^{1/3}x_2J_3}{3(x_1+\ii\, x_2)}\nn\\
\{x_2,J_2\}'&=&\dfrac{\ii\, x_3^{4/3}J_1}{x_1+\ii\, x_2}
-\dfrac{x_1J_2}{3x_3^{2/3}}-\dfrac{4\ii\, x_3^{1/3}x_1J_3}{3(x_1+\ii\, x_2)}\nn\\
\{x_1,J_2\}'&=&\dfrac{\ii\, (x_2^2-\ii\, x_1x_2-3x_3^2)J_2}{3x_3^{2/3}(x_1+\ii\, x_2)}
+\dfrac{4\ii\, x_3^{1/3}x_2J_3}{3(x_1+\ii\, x_2)}\nn\\
\{x_2,J_2\}'&=&-\dfrac{(x_1^2+\ii\, x_1x_2-3x_3^2)J_1}{3x_3^{2/3}(x_1+\ii\, x_2)}
-\dfrac{4x_3^{1/3}x_1J_3}{3(x_1+\ii\, x_2)}\nn\\
\{x_3,J_1\}'&=&-\dfrac{x_3^{1/3}(J_1x_2-x_1J_2)}{x_1+\ii\, x_2},\qquad
\{x_3,J_2\}'=-\dfrac{\ii\, x_3^{1/3}(J_1x_2-x_1J_2)}{x_1+\ii\, x_2},\nn\\
\{J_i,J_j\}'&=&\dfrac{-a\varepsilon_{ijk}x_k}{2x_3^{1/3}(x_1+\ii\, x_2)}
+\delta_{i1}\delta_{j2}\dfrac{\ii\, (J_1x_2-x_1J_2)(J_1+\ii\, J_2)}{3x_3^{2/3}(x_1+\ii\, x_2)}.\nn
\en
The corresponding Poisson bivector $P'$ is compatible with canonical Poisson bivector on $e^*(3)$
and satisfies to equations (\ref{m-eq3}), so that $P'$ has the same foliations by symplectic leaves as  $P$ at $C_2=0$.

According to \cite{fp02} the bi-involutivity of integrals of motion
\[\{H_1^{(2)},H_2^{(2)}\}=\{H_1^{(2)},H_2^{(2)}\}'=0\]
is equivalent to the existence of  non-degenerate control matrix $F$ such that
\bq
P'dH_i^{(2)}=P\sum_{j=1}^2 F_{ij}\,dH_j^{(2)},\qquad i=1,2.
\eq
In our case the control matrix $F$ reads as
\[
F=\left(
    \begin{array}{cc}
      -\dfrac{x_3^{1/3}(J_1+\ii\, J_2)}{x_1+\ii\, x_2} +\dfrac{2J_3}{3x_3^{2/3}}& \dfrac{1}{x_3^{2/3}} \\
      \\
      \dfrac{-ax_3^{2/3}}{2(x_1+\ii\, x_2)}+\dfrac{2x_3^{1/3}(J_1+\ii\, J_2)J_3}{3(x_1+\ii\, x_2)}
      -\dfrac{4J_3^2}{9x_3^{2/3}}\qquad & \dfrac{-2J_3}{3x_3^{2/3}}
    \end{array}
  \right)\,.
\]
The Darboux-Nijenhuis variables $q_{1,2}$ are simultaneously eigenvalues of the recursion operator and eigenvalues the control matrix. In our case they are the roots of the following polynomial
\[
\mathcal A(\lambda)=\det(F-\lambda\,\mathrm I)=\lambda^2+\dfrac{x_3^{1/3}(J_1+\ii\, J_2)}{x_1+\ii\, x_2}\,\lambda
+\dfrac{a}{2(x_1+\ii\, x_2)}\,.
\]
The left eigenvectors of $F$, if suitable normalized, form the St\"ackel matrix $S$, which
enters into a pair of the separated relations
\bq\label{s-rel}
\sum_{j=1}^2 S_{ij}(q_i,p_i)H_j^{(2)}-U_i(q_i,p_i)=0\,,\qquad i=1,2.
\eq
Here $U_i$ are the St\"ackel potentials and $p_{1,2}$ are variables conjugated to $q_{1,2}$:
\[
\{q_i,p_j\}=\delta_{ij},\qquad \{q_i,p_j\}'=\delta_{ij}q_i,\qquad
\{q_i,q_j\}=\{q_i,q_j\}'=\{p_i,p_j\}=\{p_i,p_j\}'=0\,.
\]
Unfortunately, construction of the variables $p_{1,2}$ is non-algorithmic procedure,
which depends on the fortune and skilfulness \cite{fp02}.  In our case we can observe that
\[
q_1+q_2=\ii\, u^{1/3}p_u-\dfrac{u^{4/3}p_\phi}{1-u^2},\qquad\mbox{and}\qquad
 q_1q_2=-\dfrac{\ii\, a\,\e^{\ii\, \phi}}{2\sqrt{1-u^2}}
\]
so that
\[
\{p_\phi,q_1+q_2\}=0,\qquad \{p_\phi,q_1q_2\}=-\ii\, q_1q_2,
\qquad
\{u,q_1+q_2\}=\ii\, u^{1/3},\qquad \{u,q_1q_2\}=0\,.
\]
Integrating these equations with respect to $p_\phi(q,p)$ and $u(q,p)$ we can easily get the following
expressions for these functions
\[
p_\phi=\dfrac{\ii\, q_1q_2(p_2-p_1)}{q_1-q_2},\qquad\mbox{and}\qquad
u=\left(\dfrac{-2\,\ii\,(q_1p_1-q_2p_2)}{3(q_1-q_2)}\right)^{3/2},
\]
which yield the necessary definitions of the momenta
\[
p_i= \mathcal B(\lambda=q_i),\quad i=1,2,\qquad\mbox{where}\qquad \mathcal B(\lambda)=\ii\,\left(\dfrac{3u^{2/3}}{2}-\dfrac{p_\phi}{\lambda}   \right)=\ii\,\left(\dfrac{3x_3^{2/3}}{2}+\dfrac{J_3}{\lambda}   \right).
\]
Using the following relations for the second Poisson bracket
\[
\{p_\phi,q_1+q_2\}'=q_1q_2,\qquad\mbox{and}\qquad
\{p_\phi,q_1q_2\}'=0
\]
it is easy to prove that $\{q_i,p_j\}'=\delta_{ij}q_i$.

In the separated variables the St\"ackel matrix $S$ is equal to
\[
S=\left(
    \begin{array}{cc}
      1 & 1 \\
      \\
      \dfrac{3\,\ii}{2q_1p_1} & \dfrac{3\,\ii}{2q_2p_2}  \\
    \end{array}
  \right)=\left(
    \begin{array}{cc}
      1 & 1 \\
      \\
      \dfrac{-1}{2\,\ii\,\alpha\,q_1p_1} & \dfrac{-1}{2\,\ii\,\alpha\,q_2p_2}  \\
    \end{array}
  \right)
\]
whereas integrals of motion have the generalized St\"ackel form at the second case (\ref{H5})
\[
H_i^{(2)}=\sum_{j=1}^{2} S^{-1}_{ij}\left(
\dfrac{(4p_j^2+6\,\ii p_j-9)(3\,\ii-2p_j)q_j^2}{18p_j}-\dfrac{3\,\ii a^2}{8q_j^2p_j}+\dfrac{3\,\ii b}{2p_j}
\right)\,,\qquad i=1,2.
\]
The corresponding separated relations (\ref{s-rel})
define two copies of the following algebraic curve
\bq\label{c1}
\mathcal C:\qquad \mu^3-H_1^{(2)}\mu+\lambda^3=b\lambda-H^{(2)}_2-\dfrac{a^2}{4\lambda},\qquad \mu=2\,\ii\,\alpha\, q_ip_i,\quad\lambda=q_i\,.
\eq
In contrast with the usual St\"ackel systems it is non-hyperelliptic curve, which is trigonal algebraic curve.
It is interesting that  one formally gets the real algebraic curve. But we have to bear firmly in mind that our separated variables $q_{1,2}$ and $p_{1,2}$ are complex functions on initial physical variables.

If we put $\nu=0$ and $\mu=1$ in $P_{\mu\nu}'$ (\ref{p-mn}) one gets another Poisson bivector on $T^*{\mathcal S}^2$:
\[
P'=\left(
     \begin{array}{cccc}
       0 & u & \ii\,up_u-\dfrac{2+u^2}{3(1-u^2)}p_\phi & 0 \\
\\
       * & 0 &\dfrac{1-4u^2}{3(1-u^2)}p_u+\dfrac{\ii\,u(4-u^2)}{3(1-u^2)^2}p_\phi  & -\dfrac{2}{3}p_\phi \\
\\
       * & * & 0 & \dfrac{\ii\,a\,u^{1/3}\e^{\ii\,\phi}}{2\sqrt{1-u^2}} \\
\\
       * & * & * & 0
     \end{array}
   \right)
\]
In this case the Darboux-Nijenhuis coordinates are roots of the following polynomial
\[
\mathcal A(\lambda)=\lambda^2-\left(
\ii\,up_u-\dfrac{4-u^2}{3(1-u^2)}p_\phi
\right)-\dfrac{\ii\,au^{4/3}\e^{\ii\phi}}{2\sqrt{1-u^2}}
-\dfrac{2\,\ii\,u}{3}p_up_\phi+\dfrac{2(2+u^2)}{9(1-u^2)}p_\phi^2\,.
\]
In term of the previous separated variables this polynomial looks like
\[
\mathcal A(\lambda)=\left(\lambda+\dfrac{2\,\ii}{3}\,q_1p_1\right)\left(\lambda+\dfrac{2\,\ii}{3}\,q_2p_2\right)
\,.
\]
Now it is easy to prove that in the separated  variables $(q,p)$ bivector $P'_{\mu\nu}$ (\ref{p-mn}) looks like
\bq\label{s-pmn}
P'_{\mu\nu}=
-2\,\ii\,\alpha\,\mu\left(
                               \begin{array}{cccc}
                                 0 & 0 & q_1p_1 & 0 \\
                                 0 & 0 & 0 & q_2p_2 \\
                                 -q_1p_1 & 0 & 0 & 0 \\
                                 0 & -q_2p_2 & 0 & 0 \\
                               \end{array}
                             \right)
-\dfrac{\nu}{a^2}\left(
                               \begin{array}{cccc}
                                 0 & 0 & q_1 & 0 \\
                                 0 & 0 & 0 & q_2 \\
                                 -q_1 & 0 & 0 & 0 \\
                                 0 & -q_2 & 0 & 0 \\
                               \end{array}
                             \right)
\eq
For the remaining four systems bivector $P_{\mu,\nu}$ (\ref{p-mn}) has the similar form in the Darboux-Nijenhuis variables $(q,p)$.

\subsection{The Goryachev-Chaplygin top}
Now we briefly discuss new separation of variables for the Goryachev-Chaplygin top \cite{ch48}. Remind, that in this case the Hamiltonian reads as
\[
H_1^{(1)}=J_1^2+J_2^2+4J_3^2-ax_1+\dfrac{b}{x_3^2}\,.
\]
The new separated coordinates are roots of the polynomial
\ben
\mathcal A(\lambda)&=&\lambda^2
+a^2\left(\dfrac{ip_u}{m n}+\dfrac{3\alpha p_\phi}{mn(u+c)}+\dfrac{p_\phi(uc+1)}{mn(u+c)(u^2-1)}\right)\lambda
+\dfrac{\ii\,a^4\,\e^{\ii\,\phi}}{2mn^2\sqrt{1-u^2}}\nn\\
\nn\\
&=&\lambda^2
-\left(\dfrac{\ii\,p_u}{u}+\dfrac{3p_\phi}{u^2}+\dfrac{p_\phi}{u^2(u^2-1)}\right)\lambda+
\dfrac{\ii\,a\,\e^{\ii\,\phi}}{2u^2\sqrt{1-u^2}}\,,\nn
\en
whereas the conjugated momenta are equal to
\[
p_i= \mathcal B(\lambda=q_i),\qquad\mbox{where}\qquad \mathcal B(\lambda)=\ii\,\left(\dfrac{u^{2}}{2}-\dfrac{p_\phi}{\lambda}   \right)\,.
\]
These separated variables lie on  two copies of the following algebraic curve
\bq\label{c2}
\mathcal C:\qquad\mu\Bigl(\mu^2+\mu\lambda-H_1^{(1)}\Bigr)=b\lambda-H_2^{(1)}-\dfrac{a^2}{4\lambda}, \qquad \mu=2\,\ii\,\alpha\, q_ip_i,\quad\lambda=q_i\,.
\eq
As for Goryachev model these separated variables $q_{1,2}$ and $p_{1,2}$ are complex functions on initial physical variables, which lie on the non-hyperelliptic algebraic curve.

Of course, we can repeat similar calculations for the remaining systems and prove that they are related with different trigonal curves as well. As an example, for the Dullin-Matveev system \cite{dull04} equation of motion are linearized on the following algebraic curve
\bq\label{c3}
\mathcal C:\qquad\mu\Bigl(\mu^2-\lambda^2- H_1^{(4)}\Bigr)=b\lambda-H_2^{(4)}-\dfrac{a^2}{4\lambda},\qquad
\mu=2\,\ii\,\alpha\, q_ip_i,\quad\lambda=q_i\,.
\eq
if $c=0$ in the Hamiltonian $H_1^{(4)}$ (\ref{H5}).

It will be interesting to get solutions of the equations of motion in term of the abelian functions for trigonal curves, as example see \cite{ee07}. The other open question is construction of the Lax matrices associated with separated variables on trigonal curves.

\subsection{The Jacobi method}
The Jacobi method consists of construction of the integrable system
starting with some known separated variables and arbitrary separated relations.
The method was originally formulated by Jacobi when
he invented elliptic coordinates and successfully applied them to solve several
important mechanical problems, such as the problem of geodesic motion on an
ellipsoid, and the problem of planar motion in a force field of two attracting
centers. In \cite{jac66} Jacobi himself wrote:
"The main difficulty in integrating a given differential equation lies in introducing convenient variables, which there is no rule for finding. Therefore, we must travel the reverse path and after finding some notable substitution, look for problems to which it can be successfully applied."

In our case this notable substitution is described by the following
\begin{prop}
At any $a,c$ and $\alpha$ transformation
\ben
u&=& \left(\dfrac{-2\,\ii\,\alpha(p_1q_1-q_2p_2)}{q_1-q_2}\right)^{1/2\alpha}-c,\qquad
\qquad
p_\phi= \dfrac{-\ii\,q_1q_2(p_1-p_2)}{q_1-q_2}\,,\label{c-tr}\\
\nn\\
\phi&=&-\dfrac{\ii}{2}\left(2\ln(u+c)(3\alpha-1)+\ln\left(\dfrac{4q_1^2q_2^2(u^2-1)}{a^2}\right)\right),\nn\\
\nn\\
p_u&=&\left(\dfrac{3\,\ii\,\alpha}{u+c} +\dfrac{\ii\,(uc+1)}{(u+c)(u^2-1)}\right)p_\phi-\ii\,(q_1+q_2)(u+c)^{2\alpha-1}\nn
\en
is canonical transformation.
\end{prop}
Now in order to get some integrable system  on $T^*\mathcal S^2$ we could take two copies of any algebraic curve $\mathcal C$ defined by equation $\Phi(\mu,\lambda,H_1,H_2)=0$ and solve the corresponding separated relations
\[\Phi(\mu_i,\lambda_i,\,H_1,H_2)=0,\qquad
\mu_i=2\,\ii\,\alpha\, q_ip_i,\quad\lambda_i=q_i
\]
with respect to integrals of motion $H_1$ and $H_2$, which will be  in bi-involution \cite{ts08b}. As above the main problem is that change of variables (\ref{c-tr}) is the transformation over the complex field $\mathbb C$ and if we want to get real functions $H_{1,2}$ on the initial variables  on $T^*\mathcal S^2$ we have to start with the very special algebraic curves $\mathcal C$, for instance with (\ref{c1},\ref{c2}) or (\ref{c3}).

\textbf{Example:} Let us consider the following deformation of the algebraic curve (\ref{c2}):
\bq\label{cá2}
\widetilde{\mathcal C}:\qquad\mu\Bigl(\mu^2+\mu\lambda-H_1\Bigr)
+\rho\left(\dfrac{\rho}{4}+\mu\right)(\lambda+\mu)
=b\lambda-H_2-\dfrac{a^2}{4\lambda}\,.
\eq
Solving the corresponding separated relations with respect to $H_{1,2}$ one gets integrals of motion in the bi-involution. After  canonical transformation (\ref{c-tr}) at $c=0$ and $\alpha=1$ we obtain the following complex Hamiltonian
\[
H_1=
(1-u^2)p_u^2+\dfrac{(4-3u^2)p_\phi^2}{1-u^2}-a\sin(\phi)\sqrt{1-u^2}+\dfrac{b}{u^2}
+\rho p_\phi-\dfrac{\rho^2(1-u^2)}{4u^2}+\dfrac{\ii\,\rho p_u(1-u^2)}{u},
\]
which after an obvious additional shift of momenta (canonical transformation)
\[
p_u\to \tilde{p}_u= {p}_u-\dfrac{\ii\,\rho}{2u}
\]
becomes the real Hamiltonian for the Goryachev-Chaplygin gyrostat
\[
H_1=H_1^{(1)}+\rho p_\phi=J_1^2+J_2^2+4J_3^2-\rho J_3-ax_1+\dfrac{b}{x_3^2}\,.
\]
More complicated deformations of the  algebraic curve (\ref{c2}), such as
\[
\mu\Bigl(\mu^2+\mu\lambda-H_1\Bigr)+
c_1\mu\lambda+c_2\lambda^2+ c_3\mu^2=b\lambda-H_2-\dfrac{a^2}{4\lambda},
\]
and the corresponding additional shifts of the momenta $p_u$ lead to 
another generalizations of the Goryachev-Chaplygin top, which was obtained in \cite{yeh02}. 

Of course, we can get similar deformations for the remaining four systems in (\ref{H5}) as well.

\section{Conclusion}
We found two-parametric Poisson bivector (\ref{p-mn}), which is compatible with canonical Poisson bivector on cotangent bundle $T^*\mathcal S^2$ of two-dimensional sphere. The quadratic and cubic integrals of motion (\ref{H},\ref{K}) for the five integrable systems on the sphere are in bi-involution with respect to the corresponding Poisson brackets.

The eigenvalues of the corresponding recursion operator are the separated coordinates. For the Goryachev system and Goryachev-Chaplygin top we give an explicit formulae for these separated variables and the corresponding separated relations.

The research was partially supported by the RFBR grant.

\end{document}